\documentclass[aps,pre,twocolumn,superscriptaddress,nobibnotes,nodoi,noeprint]{revtex4-1}
\usepackage{amsmath,amssymb,physics,bm,siunitx}
\usepackage{xcolor,graphicx}
\usepackage{soul}
\usepackage[colorlinks=true,citecolor=blue,urlcolor=blue,linkcolor=black]{hyperref}
\usepackage{setspace}

\def\rms{\rm\scriptscriptstyle}

\newcommand{\fext}{f^{\rm ext}}
\newcommand{\bfext}{\bar f^{\rm ext}}
\newcommand{\ftot}{f^{\rm tot}}

\begin{document}

\title{Solitons in overdamped Brownian dynamics}

\author{Alexander P.\ Antonov}
\email{alantonov@uos.de}
\affiliation{Universit\"{a}t Osnabr\"{u}ck, Fachbereich Physik, Barbarastra{\ss}e 7, D-49076 Osnabr\"uck, Germany}

\author{Artem Ryabov}
\email{rjabov.a@gmail.com}
\affiliation{Charles University, Faculty of Mathematics and Physics, Department of Macromolecular Physics, V Hole\v{s}ovi\v{c}k\'{a}ch 2, 
CZ-18000 Praha 8, Czech Republic}

\author{Philipp Maass} 
\email{maass@uos.de}
\affiliation{Universit\"{a}t Osnabr\"{u}ck, Fachbereich Physik, Barbarastra{\ss}e 7, D-49076 Osnabr\"uck, Germany}

\date{April 28, 2022, revised June 23, 2022} 

\begin{abstract}
Solitons are commonly known as waves that propagate without
dispersion. Here we show that they can occur for driven overdamped
Brownian dynamics of hard spheres in periodic potentials at high
densities. The solitons manifest themselves as periodic sequences of
different assemblies of particles moving in the limit of zero noise,
where transport of single particles is not possible.  They give rise
to particle currents at even low temperature that appear in band-like
structures around certain hard-sphere diameters.  At high
temperatures, the band-like structures are washed out by the noise,
but the particle transport is still dominated by the solitons. All
these predicted features should occur in a broad class of periodic
systems and are amenable to experimental tests.
\end{abstract}

\maketitle

\noindent
Brownian motion occurs ubiquitously in many natural systems and is a
relevant process in several technological applications. At high
particle densities, its properties are strongly influenced by
collective effects arising from many-body interactions. These effects
become particularly pronounced for motion in channel-like structures,
where the spatial confinement hinders particles to overtake each
other.  Examples of such single-file transport include particle motion
through membranes \cite{Hannesschlaeger/etal:2019}, nanopores
\cite{Bauer/Nadler:2006, Kahms/etal:2009, Bressloff/Newby:2013,
  Luan/Zhou:2018, Zeng/etal:2018} and in nanofluidic devices
\cite{Cheng/Bowers:2007, Dvoyashkin/etal:2014, Ma/etal:2015,
  Su/etal:2019, Ebrahimi/etal:2020, Kavokine/etal:2021}, water
permeation in nanomembranes \cite{Horner/Pohl:2018, Rezvova/etal:2020,
  Song/etal:2020, Suk:2020, Pfeffermann/etal:2021,
  Pfeffermann/etal:2022}, and catalytic processes in zeolites
\cite{Hahn/etal:1996, VanDeVoorde/Sels:2017}.  Recent experiments on
colloids driven by optical and magnetic fields allow to explore and
investigate collective effects under well-controlled conditions
\cite{Wei/etal:2000, Lutz/etal:2004, Lutz/etal:2006,
  Mirzaee-Kakhki/etal:2020, Villada-Balbuena/etal:2021,
  Cereceda-Lopez/etal:2021}.

Here we show that overdamped Brownian motion of hard spheres through
periodic channel-like geometries has a puzzling behavior under
external driving. This is reflected in a band-like structure of the
particle current in dependence of the particle diameter $\sigma$: only
in small bands around certain $\sigma$, particle transport is possible
and the current is nonzero.  We explain this behavior by well-defined
deterministic motions of local excitations in the zero-noise limit.
They represent propagating solitons of local density fluctuations.
The solitons appear if $\sigma$ lies in the bands and they govern the
current behavior in the presence of noise also. At low noise, the
band-like structure of the current is smeared out. At high noise,
particle transport becomes possible for all $\sigma$ but the magnitude
of the current is still governed by the solitons.  Changes of currents
can be traced back to the solitons even for strong driving up to the
critical tilting force, where potential barriers for particle motion
disappear. We demonstrate these findings for a simple setup, which is
amenable to direct experimental investigations.

\begin{figure}[b!]
\includegraphics[width=0.48\textwidth]{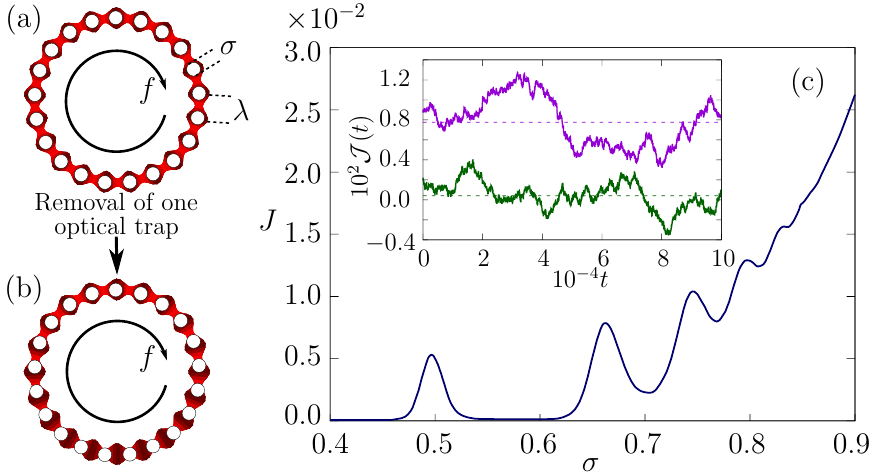}
\caption{(a) Ring of $(N+1)=21$ optical traps filled with $(N+1)$ hard
  spheres with diameter $\sigma$ in a fluid environment. The 
  shades of red indicate the value of the optical potential. 
  Areas marked in dark red (light red) correspond to potential minima (maxima).
  A drag force
  $f$ tries to drive the particles in clockwise direction but the
  system is jammed and the particle positions fluctuate around the
  points of mechanical equilibria in the traps.  (b) After removal of
  one trap (corresponding to an increase of the wavelength of the
  optical potential), particle transport becomes possible.  (c) The
  resulting stationary current $J(\sigma,f)$ shows peaks as a function
  of $\sigma$ at $\sigma_n=(n\!-\!1)/n$, $n=2,3,\ldots$ Data are shown
  for $f=0.05$ and noise strength $D=0.01$. In the inset, two
  stochastic time evolutions of the instantaneous current are
  displayed for $\sigma=\sigma_3=2/3$ (purple curve)
  and $\sigma=0.62$ (green curve). A moving
  average of the instantaneous current was taken for a time window of
  size $2.5\times10^4$ to reduce the noise level.}
\label{fig:model}
\end{figure}

The setup consists of a ring of 21 optical traps in a fluidic
environment, where each trap is occupied by one particle, as
illustrated in Fig.~\ref{fig:model}(a).  A constant drag force tries
to move the particles in clockwise direction, but there is no motion
as the system is jammed.  At some time instant, one well is removed,
which leads to a ``distorted particle configuration'', where a regular
filling of each well by one particle is no longer possible, see
Fig.~\ref{fig:model}(b). We consider the subsequent Brownian particle
dynamics given by the Langevin equations
\begin{equation}
\frac{\dd x_i}{\dd t}=\mu\left[f -U'(x_i)\right] + \sqrt{2D}\,\xi_i(t)\,,\hspace{0.5em}i=1,\ldots,N\!+\!1\,,
\label{eq:langevin}
\end{equation}
where $U(x)=(U_1/2)\cos(2\pi x/\lambda)$ describes the periodic
optical potential, $f$ is the constant drag force, $\mu$ is the
particle mobility, $D=k_{\rms{B}}T\mu$ is the diffusion coefficient
with $k_{\rms B}T$ the thermal energy. The $\xi_i(t)$ are Gaussian
white noise processes with zero mean and correlations $\langle
\xi_i(t) \xi_j(t') \rangle = \delta_{ij}\delta(t - t')$. The distance
between neighboring particles cannot be smaller than $\sigma$
(hard-sphere interaction) and the particles keep their order
(single-file motion).  This system represents a minimal model for
studying Brownian single-file transport in periodic potentials
\cite{Lips/etal:2018, Ryabov/etal:2019, Lips/etal:2019,
  Cereceda-Lopez/etal:2021}. We set $\mu=1$ in the following and
choose $\lambda$, $\lambda^2/\mu U_1$ and $U_1$ as units for length,
time and energy, respectively.

The hardcore interactions do not enter the equations of motion~\eqref{eq:langevin} explicitly due to the
singular form of the hard-sphere potential. The interactions are taken into account by the constraint that 
neighboring particles must have a distance larger than or equal to $\sigma$. 
Our procedure to generate particle trajectories 
is described in the supplemental
material (SM), see below.

Using this procedure, we determined currents in
the nonequilibrium steady state.  This state forms after a short
transient time after the removal of one trap.  The instantaneous
current is $\mathcal{J}(t;\sigma,f)=(N+1)\bar v(t)/L$ with $\bar v(t)$
being the center of mass velocity, $\bar v(t)=[\sum_{j=1}^{N+1} \dd
  x_j(t)/\dd t]/(N\!+\!1)$. After averaging $\mathcal{J}(t;\sigma,f)$
over time, we obtain the stationary current $J(\sigma,f)$.

Figure~\ref{fig:model}(c) shows $J(\sigma,f)$ as a function of
$\sigma$ for small driving $f=0.05$ and low noise $D=0.01$ ($k_{\rm
  B}T/U_1=0.01$), where a single particle would hardly surmount
potential barriers.  For $\sigma<0.4$ (not shown) the system is indeed
essentially jammed, i.e.\ there is no noticeable current. However,
$J(\sigma,f)$ is not always negligibly small.  Besides nearly jammed
states , phases of ``running states'' occur, where significant
particle transport is present. The current shows peaks around maxima
at particle diameters
\begin{equation}
\sigma_n=\frac{n-1}{n}\,,\hspace{1em} n=2,3,\ldots\,,
\label{eq:sigman}
\end{equation}
where the maxima $J(\sigma_n,f\!=\!0.05)$ in Fig.~\ref{fig:model}(c)
increase with increasing $n$.  Between the peak positions at
$\sigma_2=1/2$ and $\sigma_3=2/3$, the current drops to small values.
As the distances between neighboring peaks become smaller with
increasing $n$, the peaks overlap and the current runs through minima
with values significantly larger than zero.

\begin{figure}[t!]
\includegraphics[width=0.48\textwidth]{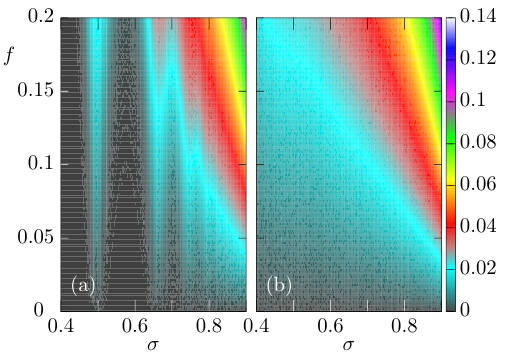}
\caption{Current $J(\sigma,f)$ as a function of $\sigma$ and $f$ in a
  color-coded representation for different noise strengths
  (temperatures) (a) $D=0.01$ and (b) $D=0.1$.}
\label{fig:j-f-sigma}
\end{figure}

Figure~\ref{fig:j-f-sigma} shows how the system changes between jammed
and running states in dependence of $\sigma$ and $f$ at two
temperatures (or noise strengths $D$) in the regime of weak driving
($0<f\le0.2$). The jammed states refer to regions, where $J(\sigma,f)$
is vanishingly small (black areas in the figure), and the running
states to regions with noticeable $J(\sigma,f)$, as indicated by the
color coding (cf.\ scale bar in the figure).

For low noise $D=0.01$ [Fig.~\ref{fig:j-f-sigma}(a)], bands of running
states can be seen around the $\sigma_n$ given in
Eq.~\eqref{eq:sigman}.  When increasing $f$, the bands widen. We can
furthermore identify lines of local maxima in the current, see the
lines of highest color brightness in
Fig.~\ref{fig:j-f-sigma}(a). These lines appear to be centered in
stripes of the same color which form a pattern in the upper right half
of Fig.~\ref{fig:j-f-sigma}(a). For the larger noise $D=0.1$
[Fig.~\ref{fig:j-f-sigma}(b)], the bands around the $\sigma_n$ can no
longer be seen but the stripe-like pattern is still present.

How can we understand the occurrence of the alternating phases of
jammed and running states? Since we obtain currents even for low
noise, let us consider the transport behavior in the limit of zero
noise ($D=0$). In this case, the equations of motion
\eqref{eq:langevin} become deterministic.  Individual particles cannot
surmount any barrier and one may wonder, how particle transport is
possible without thermal activation. Collective effects must play a
role. They manifest themselves in the formation of $n$-clusters, which
consist of $n$ mutually touching particles, i.e.\ where the distance
between nearest neighboring particles equals $\sigma$.

The effective potential $U_n(x)=\sum_{j=1}^n U(x+(j\!-\!1)\sigma)$ for
an $n$-cluster to move in the cosine potential $U(x)$ is
\begin{equation}
U_n(x)=\frac{U_1}{2}\,\frac{\sin(n\pi\sigma)}{\sin(\pi\sigma)}\cos[2\pi x+(n-1)\pi\sigma]\,.
\label{eq:Un}
\end{equation}
It has the form as that for a single particle but with a modified
potential barrier $U_1\sin(n\pi\sigma)/\sin(\pi\sigma)$ that vanishes
for the $\sigma_n$ in Eq.~\eqref{eq:sigman}. Accordingly, if $n$
particles of size $\sigma_n$ would form a moving cluster, their
collective motion requires no thermal activation. This could be a
possible explanation for the occurrence of peaks in the current at
$\sigma=\sigma_n$.

\begin{figure*}[t!]
\includegraphics[width=\textwidth]{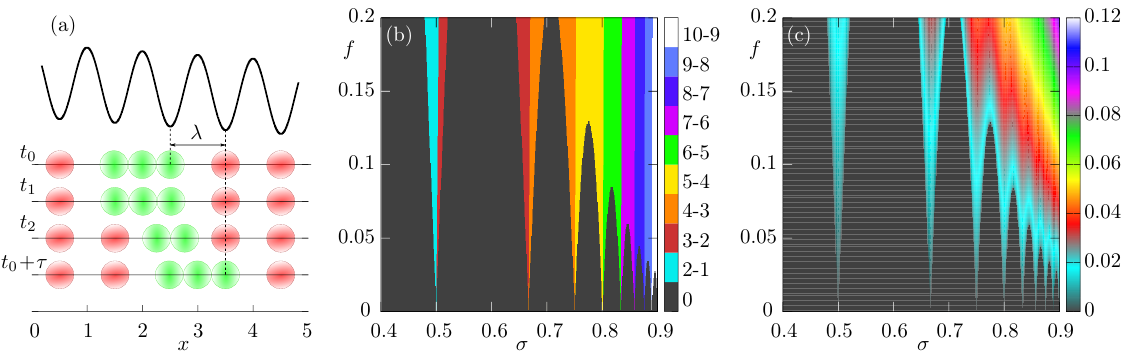}
\caption{Solitons and their impact on the current in the zero-noise
  limit. (a) Propagation of a 3-2-soliton for $f=0.2$ and
  $\sigma=0.51$. A 3-cluster (green particles) starts to move at time
  $t_0$. Its leftmost particle detaches at time $t_1 = t_0 + 0.102$.
  At that time instant, the rightmost particle of the cluster 
  would not be able to further move as a single particle, because
 it had to overcome a barrier, 
 see the tilted cosine potential at the top of the graph.
  However, due its contact
 with a neighboring particle to the left, the total force on the cluster formed by the two particles
 is positive. Accordingly, the 2-cluster continues to move (time
  $t_2=t_0+1.906$). Eventually, the 2-cluster is attaching to a
  resting particle after the period $\tau=2.764$, where the process
  starts again.  Particles marked in green belong to propagating
  clusters, and the particles marked in red are relaxing to positions
  of mechanical equilibria (being very close to them in the steady
  state). (b) Occurrence of $n$-$(n\!-\!1)$ solitons and (c)
  stationary current $J(\sigma,f)$ in dependence of $\sigma$ and weak
  driving forces $f\le0.2$.}
\label{fig:zero_noise}
\end{figure*}

But can $n$ particles remain attached? Actually, as we show now, this
not the case. The cluster dynamics is more complex and given by
certain conditions on the forces
\begin{equation}
F_i=F(x_i)=f-U'(x_i)
\label{eq:F}
\end{equation}
acting on the particles.  Let us first consider a 2-cluster with
particles at positions $x_1$ and $x_2=x_1+\sigma$. If $F_1$ is
positive and $F_2$ smaller than $F_1$, the two particles must move
together, i.e.\ the 2-cluster moves as a whole. If $F_1$ is negative,
one can draw an analogous conclusion, yielding the general condition
$F_1\ge F_2$ for the 2-cluster to move as a whole. If $F_1<F_2$ by
contrast, the particles detach.

For $n$-clusters with $n>2$, the possibilities of movements become
richer. For example, a 3-cluster has the $2^2=4$ compositions
$\{111\}$, $\{2,1\}$, $\{1,2\}$, and $\{3\}$, which specify the
possible movements: In the composition $\{111\}$, all particles become
detached. In $\{2,1\}$, the first two particles move together,
i.e.\ they form a subcluster, and the third particle detaches from the
second. In $\{1,2\}$, the last two particles move together as a
subcluster and detach from the first, and in composition $\{3\}$ all
three particles keep in touch and the 3-cluster moves as a whole.

Considering a general composition $\{m_1\ldots m_s\}$ of a cluster of
size $n=\sum_{j=1}^sm_j$, the subclusters $k$ of size $m_k$ have the
velocity $\mu\bar F_k$ ($k=1,\ldots,s$), where $\bar F_k$ is the mean
force on the subcluster $k$. As each subcluster must detach from its
neighboring ones, one set of conditions for an $n$-cluster to move
according to the composition $\{m_1\ldots m_s\}$ is
\begin{subequations}
\begin{equation}
\bar F_k<\bar F_{k+1}\,,\hspace{1em}k=1,\ldots,s-1\,.
\label{eq:cond1}
\end{equation}
This is the analogue to the situation of two particles considered
above.  In addition, particles within a subcluster must not detach,
i.e.\ for all possible divisions of a subcluster into two
sub-subclusters, the inequality \eqref{eq:cond1} must be violated.
This implies that for each subcluster $k=1,\ldots,s$ of size $m_k$ in
the composition $\{m_1\ldots m_s\}$, it must hold
\begin{equation}
\frac{1}{i}\sum_{j=1}^iF_{k,j}\ge \frac{1}{m_k-i}\sum_{j=i+1}^{m_k} F_{k,j}\,, \hspace{1em}i=1,\ldots,m_k-1\,,
\label{eq:cond2}
\end{equation}
\end{subequations}
where $F_{k,j}=F_l$ is the force on particle $j$ in the $k$-th
subcluster ($l=m_1+m_2+\ldots m_{k-1}+j$). Note that the conditions
\eqref{eq:cond1} and \eqref{eq:cond2} are independent of the drag
force $f$.  A more detailed derivation of them is given in SM.

Solving Eqs.~\eqref{eq:langevin} in the zero-noise limit subject to
the conditions \eqref{eq:cond1} and \eqref{eq:cond2}, we find specific
types of soliton-like movements that occur for $\sigma=\sigma_n$ and
$\sigma$ close to $\sigma_n$ from Eq.~\eqref{eq:sigman} for larger
$f$. An example of such soliton is shown for $\sigma=0.51$ and $f=0.2$
in Fig.~\ref{fig:zero_noise}(a). At an initial time $t_0$, a 3-cluster
starts moving, where its last particle is at a position of mechanical
equilibrium, i.e.\ a minimum of the tilted potential $U(x)-fx$.  This
3-cluster moves until its first particle detaches at time $t_1$.
After time $t_1$, the first particle relaxes to a position of
mechanical equilibrium, and the 2-cluster continues moving, see the
configuration at the time $t_2$ in Fig.~\ref{fig:zero_noise}(a).
Finally, the 2-cluster attaches to the particle at the position of
mechanical equilibrium right to it, leading to a new propagating
3-cluster at a time $t_0+\tau$ \footnote{Strictly speaking, in the
  steady state, all single particles are relaxing toward positions of
  mechanical equilibria (the relaxation would need infinite time in
  the zero-noise limit, if there would be no soliton). This means that
  the 2-cluster does not attach to a resting particle exactly at a
  position of mechanical equilibrium, but at a very small (negligible)
  distance away from this position.}. The process then repeats itself,
meaning that the soliton motion is periodic with $\tau$.  After one
period, the soliton has moved by one wavelength, as indicated by the
dashed lines in Fig.~\ref{fig:zero_noise}(a).  
Further examples of soliton propagation at weak driving are given in movie file~1 of
the ancillary files to this manuscript. There we demonstrate also that the
soliton propagation is essentially unaltered in the presence of weak
thermal noise. The noise leads to a rattling of particle positions
around the ones given by the deterministic time evolution.

In general, a soliton consists of a propagation composed of a periodic
sequence of cluster movements. Analogous types of solitons occur close
to all $\sigma_n$: an $n$-cluster starts to move, followed by an
$(n-1)$-cluster until a period is finished.  The emergence of these
different types of $n$-$(n\!-\!1)$-solitons in dependence of $\sigma$
and $f$ is shown in Fig.~\ref{fig:zero_noise}(b).  A band-like
structure of the solitons occurs, which is reflecting the band-like
structure of the current in the zero-noise limit shown in
Fig.~\ref{fig:zero_noise}(c), as well as in the presence of weak noise
in Fig.~\ref{fig:j-f-sigma}(a). The fact that the solitons give rise
to currents for vanishing noise means that their motion does not
require thermal activation. For the time intervals of a propagating
$n$-$(n\!-\!1)$-soliton, where an $n$-cluster moves, we know from
Eq.~\eqref{eq:Un} that no potential barrier can stop the motion. When
the $(n\!-\!1)$-cluster moves, it always runs downhill the tilted
potential.

The current generated by a soliton is $J(\sigma,f)=\bar v_{\rm
  sol}(\sigma,f)/L$, where $\bar v_{\rm
  sol}(\sigma,f)=\lambda/\tau(\sigma,f)$ is the mean velocity of a
soliton. This velocity depends on the soliton type and can be
calculated from the deterministic equation of motion, see SM.

For particle diameters $\sigma=\sigma_n$, we can estimate the current
$J(\sigma_n,f)$ by assuming that an $n$-cluster is propagating the
whole time. This is because during one period, the time intervals for
motions of clusters of other sizes are generally much smaller than
that of the $n$-cluster. For example, the soliton illustrated in
Fig.~\ref{fig:zero_noise}(a) has $\sigma$ close to $\sigma_2=1/2$ and
the 2-cluster is moving most of the time.  An $n$-cluster formed by
particles of size $\sigma_n$ does not need to surmount barriers
[$U_n=0$ in Eq.~\eqref{eq:Un}].  Accordingly, its particles propagate
with velocity $\mu f$. One period of the soliton covers the time,
where the rightmost particle of the $n$-cluster starts to move and
attaches to the next resting particle.  This means that the rightmost
particles moves a distance $(\lambda-\sigma_n)$ during one period. We
thus estimate $\tau(\sigma_n,f)\simeq(\lambda-\sigma_n)/\mu f$,
yielding
\begin{equation}
J(\sigma_n,f)=\frac{\bar v_{\rm sol}(\sigma_n,f)}{L}=\frac{\lambda}{L\tau(\sigma_n,f)}
\simeq \frac{\mu f}{N(\lambda-\sigma_n)}\,.
\label{eq:peak-values-J}
\end{equation}
Revisiting the peaks in the current in Fig.~\ref{fig:model}(c) for low
noise and $f=0.05$, the estimate \eqref{eq:peak-values-J} yields
$J(\sigma_n,0.05)\simeq 5\times 10^{-3}$, $7.5\times 10^{-3}$,
$10^{-2}$ for $n=2,3,4,\ldots$ in good agreement with the simulated
results.

The bands in Fig.~\ref{fig:zero_noise}(c) widen with increasing $f$,
because barriers become smaller and non-thermally activated motion of
solitons becomes possible for particle sizes further away from the
$\sigma_n$. Movie file~2 in the ancillary files give examples
for soliton motions at large $f$.

\begin{figure}[t!]
\includegraphics[width=0.48\textwidth]{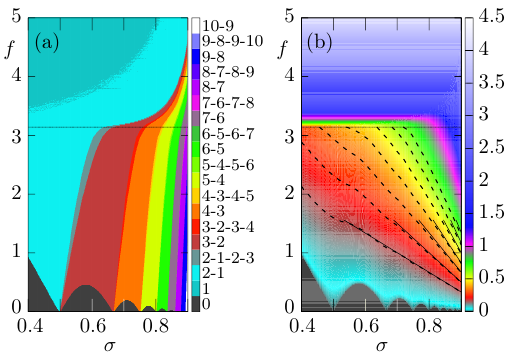}
\caption{(a) Occurrence and type of solitons in dependence of $\sigma$
  and $f$ for zero noise and extended range of driving up to and
  beyond the critical tilting force $f_c$. (b) Color-coded
  representation of currents $J(\sigma,f)$ in the zero-noise limit for
  the extended $f$-range. The dashed lines indicate isolines of
  constant currents $J_k=(k+1)/12$ for $k=1,2,\ldots,7$, and the solid
  lines represent the interpolation formula $\mu
  f/N(\lambda\!-\!\sigma)=J_k$. The horizontal dotted lines mark $f_c$.}
\label{fig:large-f}
\end{figure}

Let us see, what happens if we consider stronger driving up to the
critical tilting force $f_c=\pi$, where there would be no barriers for
a single particle, i.e.\ where $F(x)$ in Eq.~\eqref{eq:F} starts to
become positive for all $x$.  Figure~\ref{fig:large-f}(a) shows that
for this larger range of driving forces a further type of
$n$-$(n\!-\!1)$-$n$-$(n\!+\!1)$-soliton occurs involving clusters of
size $(n\!-\!1)$, $n$, and $(n\!+\!1)$ ($n=2,3,\ldots$).  The regions
of occurrence of these solitons separate those of the $n$-$(n\!-\!1)$-
and $(n\!+\!1)$-$n$-solitons. They become smaller with decreasing $f$
and terminate at the $\sigma_n$ for $f\to0$ (with extension zero). In
fact, they are so narrow for small $f$, that they cannot be resolved
in Fig.~\ref{fig:zero_noise}(b). The time interval of the movement of
the $(n\!-\!1)$-cluster is practically negligible in comparison with
the time intervals of the $n$-cluster motion, i.e.\ our estimate of
the peak currents in Eq.~\eqref{eq:peak-values-J} remains valid.

As long as $f<f_c$, the current is still governed by the solitons.
This is evident from the characteristic pattern of current values in
Fig.~\ref{fig:large-f}(b). For $f\lesssim 1.5$, the isolines of
constant currents in the $\sigma$-$f$-plane (dashed lines) follow
closely the equation $\mu f/N(\lambda-\sigma)=\textrm{const.}$ This
formula for isolines results when interpolating the peaks
$J(\sigma_n,f)$ in Eq.~\eqref{eq:peak-values-J} by replacing
$\sigma_n$ with $\sigma$.  Analogous isolines can be drawn in
Fig.~\ref{fig:j-f-sigma}, where we analyzed the impact of noise on the
soliton-induced current.  For overcritical tilting $f>f_c$, isolines
of constant current do not reflect soliton propagations, but are
almost independent of $\sigma$.  Already for a tilting slightly above
the overcritical one, the current becomes close to that of independent
particles (except for large $\sigma$).

In summary, we have shown that for driven overdamped Brownian motion
through a periodic potential, particle transport in dense systems is
possible although single particles cannot surmount the potential
barriers.  The particle current is mediated by solitons, which are
periodic sequences of particle clusters that propagate in the
zero-noise limit.  At weak driving force $f$, solitons occur only for
certain intervals of hard-sphere diameters centered around $\sigma_n$
[Eq.~\eqref{eq:sigman}].  They lead to pronounced peaks in currents
$J(\sigma,f)$ as a function of $\sigma$ for weak $f$. In the
$\sigma$-$f$-plane, the peaks correspond to bands that widen with
increasing $f$ and overlap each other at larger $f$. Soliton-governed
currents are observed up to the critical tilting force, beyond which
single-particle motion becomes possible in the zero-noise limit. The
behavior of the currents in the presence of noise (finite
temperatures) is dominated by the soliton-induced transport. For weak
noise (low temperatures), the band-like structure of the currents in
the zero-noise limit remains visible. For larger noise (higher
temperatures), the band-like structure is washed out, but the
dependence of the current $J(\sigma,f)$ on $\sigma$ and $f$ reflects
the soliton dynamics.

The solitons always occur in a crowded system independent of the number of optical traps. If the filling factor $N/M$ ($N$ is number of particles, $M$ 
number of traps) is smaller than one, they appear as rare events due to thermal fluctuations. In that case they have a finite lifetime, because they can annihilate
with empty traps. Movie file~3 of SM demonstrates a short-lived soliton in a system of 19 particles and 20 traps.
For filling factors $N/M>1$, solitons could also occur as rare events, when $\sigma$ is very different from 
$\sigma_n$ given in Eq.~\eqref{eq:sigman} (black regions in Figs.~\ref{fig:j-f-sigma}-\ref{fig:large-f}). 
For particle sizes close to $\sigma_n$,
they are permanently present then.
There are even other possible particle diameters for permanent soliton propagation in accordance with the conditions derived in SM. 
As an example, we show in movie file~4 of SM solitons for different numbers $M$ of traps and particle number $N=M+1$,
and in movie file~5 a soliton for $N=22$ particles and $M=20$ traps.

For other types of periodic potentials than the sinusoidal one considered here, solitons will occur also.
This is because the
potential for a particle cluster has the same periodicity but its barriers can
be much lower than that for a single particle, or even are
vanishing. For an arbitrary periodic potential, conditions for an
$n$-cluster to move barrier-free are derived in SM.

Traditionally, in systems with inertia, like the prominent
Frenkel-Kontorova model \cite{Braun/Kivshar:2004}, solitons are waves
whose dispersion is suppressed by nonlinear effects. In our case, the
solitons occur in the absence of inertia for fully overdamped
dynamics, where the particles keep together in the clusters because
the external forces are not able to separate them.

The general occurrence of solitons in overdamped Brownian dynamics
suggests that they can be detected in different experimental setups in
addition to the one sketched in Fig.~\ref{fig:model}.  We expect them
to play an important role in transport processes in crowded biological
systems and microfluidic devices. In the latter case, in particular,
noise is typically low and solitons can lead to significant
enhancement of currents even at high density, where often jamming
mitigates motion.

In our analysis here, we focused on hardcore interactions. For hard spheres with 
additional attractive interactions, cluster formations of particles will be easier, and
solitons should occur for wider ranges of particle diameters at given temperature and
forcing. An interesting question is, whether solitons can exist also in overdamped Brownian motion
of particles with softcore repulsive interactions. Driven particle transport for such interactions
can reflect features of hardcore systems \cite{Antonov/etal:2021}, and we
thus believe that periodic collective motions of localized particle assemblies are possible.
These and other questions, e.g.\ on likewise soliton propagation in higher dimensions and for
underdamped Brownian dynamics, open up promising perspectives for further research.

\acknowledgements{We thank D.~Lips for many valuable discussions and for
developing programs to simulate Brownian dynamics of hard spheres. 
Financial support by the Czech Science Foundation
  (Project No.\ 20-24748J) and the Deutsche Forschungsgemeinschaft
  (Project No.\ 432123484) is gratefully acknowledged.}


%


\onecolumngrid
\newpage
\renewcommand{\theequation}{S\arabic{equation}}
\renewcommand{\thefigure}{S\arabic{figure}}
\setcounter{equation}{0}
\setcounter{figure}{0}

\begin{center}
\setcounter{page}{1}
{\large\bf Supplemental Material for}\\[2ex]
{\large\bf Solitons in overdamped Brownian dynamics}\\[2ex]
Alexander Antonov,$^1$ Artem Ryabov,$^2$ and Philipp Maass,$^3$\\[2ex]
$^1$\textit{Universit\"{a}t Osnabr\"{u}ck, Fachbereich Physik, Barbarastra{\ss}e 7, D-49076 Osnabr\"uck, Germany}\\[1ex]
$^2$\textit{Charles University, Faculty of Mathematics and Physics, Department of Macromolecular Physics,\\ V Hole\v{s}ovi\v{c}k\'{a}ch 2, 
CZ-18000 Praha 8, Czech Republic}
\end{center}

\setstretch{1.5}
\vspace{1ex}\noindent

In Sec.~\ref{sec:zero-noise-dynamics} of this Supplemental Material,
we discuss how Brownian motion of hard spheres in an arbitrary
external force field can be treated in the zero-noise limit.
Conditions for barrier-free propagation of $n$-clusters in general
periodic potentials are derived in
Sec.~\ref{sec:conditions-periodic-potentials}. Their meaning for the
particle size is illustrated for a piecewise linear potential, and
potentials represented by a finite number of Fourier coefficients.  In
Sec.~\ref{sec:soliton-velocities}, we show how soliton velocities can
be calculated analytically.


\section{Overdamped Brownian motion of hard spheres in the zero-noise limit}
\label{sec:zero-noise-dynamics}
The overdamped Brownian motion of hard spheres in an arbitrary
external force field $\fext(x)$ is described by the Langevin equations
(1) in the main text. There, $\fext(x)=f-U'(x)$ with $U(x)$ being the
cosine potential. If a particle $i$ is not in contact with other
particles, its time-evolution in the zero-noise limit is
\begin{equation}
\dot x_i=\frac{\dd x_i}{\dd t}=\mu \fext(x_i)\,,
\label{eq:dxdt}
\end{equation}
where we set $\mu=1$ in the following. 

For a particle in contact with other particles, we need to take into
account interaction forces in Eq.~\eqref{eq:dxdt}.  An $n$-cluster is
formed by $n$ particles that are mutually in contact, but not in
contact with other particles.  Let $x_1$ be the coordinate of the
first particle in the cluster, and $x_i=x_1+(i-1)\sigma$,
$i=2,\ldots,n$ the coordinates of the other $(n\!-\!1)$ particles. We
pose the following question: Under which conditions do particles $i$
and $(i\!+\!1)$ in the $n$-cluster interact and what is the
interaction force then?  Knowing the answer to this question, we can
include the respective interaction forces in Eq.~\eqref{eq:dxdt} to
evolve the coordinates of particles in clusters.

The solution of the problem rests on the following properties:
\begin{list}{}{\setlength{\leftmargin}{1em}\setlength{\rightmargin}{0em}
\setlength{\itemsep}{0ex}\setlength{\topsep}{0ex}} 

\item[(i)] The interaction force $f_{i,i+1}$ of particle $i$ on
  particle $(i\!+\!1)$ in the $n$-cluster must be non-negative,
  $f_{i,i+1}\ge0$.

\item[(ii)] The interaction forces obey Newton's principle of action
  and reaction: The interaction force $f_{i+1,i}$ of particle
  $(i\!+\!1)$ on particle $i$ is $f_{i+1,i}=-f_{i,i+1}$.

\end{list}
The equations of motions for the particles in the $n$-cluster are
\vspace{-1ex}
\begin{subequations}
\label{eq:motions}
\begin{align}
\dot x_1&=\fext_1-f_{12}=\ftot_1\,,\label{eq:motions-a}\\
\dot x_i&=\fext_i+f_{i-1,i}-f_{i,i+1}=\ftot_i\,,\hspace{1em} i=2,\ldots,n\!-\!1\,,\label{eq:motions-b}\\
\dot x_n&=\fext_n+f_{n-1,n}=\ftot_n\,,\label{eq:motions-c}
\end{align}
\end{subequations}
where $\fext_i=\fext(x_i)$ and $\ftot_i$ is the total force acting on particle $i$.

The $n$-cluster can move as a whole or fragment into separate
subclusters, including 1-subclusters, i.e.\ single particles. There are
$2^{n-1}$ possible compositions $\{m_1\ldots m_s\}$, $1\le m_j\le n$,
$\sum_{j=1}^s m_j=n$, of the $n$-cluster by $m_j$-subclusters,
$s=1,\ldots,n$.

Let us consider an $m$-subcluster with particle coordinates $y_i$,
$i=1,\ldots,m$.  For the subcluster $k$ in the composition
$\{m_1\ldots m_s\}$ of the $n$-cluster, $m=m_k$ and
$y_i=x_l+(i-1)\sigma$ with $l=1+\sum_{j=1}^{k-1} m_j$.  As the
particles in the $m$-subcluster remain in contact, it must hold $\dot
y_1=\ldots=\dot y_m$.  This is caused by the interaction forces
between neighboring particles in the subcluster (unless the external
forces acting on them are all equal).  Because the interaction forces
obey the principle of action and reaction, the subcluster velocity is
given by the mean $\bfext$ of the external forces acting on the
particles, i.e.\ it holds
\begin{equation}
\dot y_1=\ldots=\dot y_m=\bfext=\frac{1}{m}\,(\fext_1+\ldots+\fext_m)\,,
\end{equation}
or $\ftot_j=\bfext$, $j=1,\ldots,m$  [$\fext_i=\fext(y_i)$ here].
Using the equations of motions for $\dot y_i$ in Eqs.~\eqref{eq:motions-a}-\eqref{eq:motions-c} 
(for $n=m$ and $x_i$ replaced by $y_i$), we obtain $m$ 
linear equations for determining the $(m\!-\!1)$ interaction forces $f_{12},f_{23},\ldots,f_{m-1,m}$. 
Only $(m\!-\!1)$ of these equations are independent, since $\sum_{i=1}^m \dot y_i=m\bfext$ does not depend on the interaction forces. 

The system of linear equations has the solution
\begin{align}
f_{i,i+1}=\sum_{j=1}^i\fext_j-i\bfext
=\frac{m\!-\!i}{m}\sum_{j=1}^i\fext_j-\frac{i}{m}\sum_{j=i+1}^m\fext_j\,,\hspace{1em}i=1,\ldots,m-1\,.
\end{align}
The interaction forces must be non-negative, which yields the
conditions
\begin{equation}
\frac{1}{i}\sum_{j=1}^i\fext_j\ge \frac{1}{m-i}\sum_{j=i+1}^m\fext_j\,,\hspace{1em}i=1,\ldots,m-1\,,
\label{eq:cond-nofragmentation-1}
\end{equation}
for the considered $m$-subcluster to move without fragmentation. These
conditions have a simple interpretation: the mean velocity of any
group of particles on the left side of the subcluster must be larger
than the mean velocity of the complimentary group of particles on the
right side, i.e.\ no group of particles on the right side can outrun
the group on the left side. Note that this is ensured by the
inequalities \eqref{eq:cond-nofragmentation-1} irrespective of the
sign of velocities (mean forces).

In addition, the considered subcluster must move independently from
the other subclusters. This requires its velocity $\bar f^{\rm ext}$
to be larger than the velocity $\bar f^{\rm ext}_-$ of its neighboring
subcluster to the left, and to be smaller than the velocity $\bar
f^{\rm ext}_+$ of its neighboring subcluster to the right,
\begin{equation}
\bar f^{\rm ext}_-<\bar f^{\rm ext}<\bar f^{\rm ext}_+\,.
\label{eq:cond-fragmentation-2}
\end{equation}
The conditions \eqref{eq:cond-nofragmentation-1} and
\eqref{eq:cond-fragmentation-2} on the external forces must be
fulfilled between all subclusters in a composition $\{m_1\ldots
m_s\}$, $1\le m_j\le n$. For the first and last subcluster of size
$m_1$ and $m_s$, there is no condition with respect to a neighboring
subcluster to the left and right, respectively.

In summary, we find that an $n$-cluster evolves into the composition $\{m_1\ldots m_s\}$, if
\begin{subequations}
\begin{gather}
\bfext_k=
\frac{1}{m_k}\sum_{j=1}^{m_k} \fext_{k,j}<\frac{1}{m_{k+1}}\sum_{j=1}^{m_{k+1}} \fext_{k+1,j}
=\bfext_{k+1}\,,
\hspace{1em}k=1,\ldots,s-1\,,
\label{eq:cond-fragmentation}\\
\frac{1}{i}\sum_{j=1}^i \fext_{k,j}\ge\frac{1}{m_k-i}\sum_{j=i+1}^{m_k} \fext_{k,j}\,, 
\hspace{1em}i=1,\ldots,m_k-1\,,\hspace{1em} k=1,\ldots,s\,,
\label{eq:cond-nofragmentation}
\end{gather}
\end{subequations}
where $\fext_{k,i}$ is the external force on particle $i$ in the
$k$-th subcluster [or the force $\fext_l$ on particle $l$ in the
  $n$-cluster, where $l=i+\sum_{j=1}^{k-1} m_j$].  The conditions
\eqref{eq:cond-fragmentation} and \eqref{eq:cond-nofragmentation}
correspond to the conditions (5a) and (5b) in the main text.

The inequalities~\eqref{eq:cond-fragmentation} and
\eqref{eq:cond-nofragmentation} give in total
$[\sum_{k=1}^s(m_k-1)+(s-1)]=(\sum_{k=1}^sm_k-1)=n-1$ conditions for a
certain composition $\{m_1\ldots m_s\}$ to occur. They are fulfilled
only for one of the possible $2^{n-1}$ compositions. Given that
composition, the time evolution of the particles in the $n$-cluster is
uniquely determined.

We finally notice that this method of evolving hard-sphere systems can
be applied also for overdamped Brownian dynamics in the presence of
noise. One simply needs to consider the random forces mediated by the
fluid as additional external forces.  We have used already the
corresponding simulation method when generating particle trajectories and compared it
to existing schemes \cite{Scala:2012, Behringer/Eichhorn:2012}.
Our simulation method can be extended to hard spheres with additional
contact interaction and will be presented elsewhere
\cite{Antonov/etal:2022-tobepublished}.

\section{Barrier-free cluster propagation in periodic potentials}
\label{sec:conditions-periodic-potentials}
Let $V(x)$ be a $\lambda$-periodic potential for single particles,
\begin{equation} 
V(x)=V(x+\lambda)\,.
\label{eq:V-single}
\end{equation}
We are interested in the behavior of the corresponding potential 
$V_n(x)$ for $n$-clusters, 
\begin{equation}
V_n(x) = \sum_{j=1}^n V(x+(j-1)\sigma)\,. 
\label{eq:Vn-cluster}
\end{equation} 
This potential is also $\lambda$-periodic and can have significantly
reduced or even vanishing barriers for certain particle diameters
$\sigma$. For these $\sigma$ values, one would obtain barrier-free
motion of an $n$-cluster composed of $n$ ``glued'' particles in mutual
contact.  If the external forces acting on the $n$-cluster satisfy
inequalities (S7a) and (S7b) over an extended part of the period
length $\lambda$, the cluster can give rise to a propagating
soliton. We here derive conditions for the barrier-free motion of
clusters.  Following the choice of length unit in the main text, we
set $\lambda=1$.

The Fourier series expansion of the periodic potential is
\begin{equation}
V(x) = \sum_{k=-\infty}^{\infty} c_k e^{2\pi i k x}
=c_0+2\Re\left[\sum_{k=1}^{\infty} c_k e^{2\pi i k x}\right]\,, 
\label{eq:V-Fourier-expansion}
\end{equation}
where 
\begin{equation}
c_k=\int_0^1\hspace{-0.2em}\dd x\, V(x)\, e^{-2\pi ikx}
\label{eq:V-Fourier-coefficients}
\end{equation}
are the Fourier coefficients.
Inserting Eq.~\eqref{eq:V-Fourier-expansion} into Eq.~\eqref{eq:Vn-cluster}, 
we obtain the Fourier series of $V_n(x)$,
\begin{align} 
V_n(x)&=nc_0+2\Re\left[\sum_{k=1}^\infty c_k e^{2\pi i k x}\sum_{j=1}^n e^{2\pi i k j \sigma}\right]
=nc_0+2\Re\left[\sum_{k=1}^\infty \frac{1-e^{2\pi ikn\sigma}}{1-e^{2\pi ik\sigma}}c_k e^{2\pi i k x}\right]\,.
\label{eq:Vn-Fourier}
\end{align}
The potential becomes constant if the factor $(1-e^{2\pi
  ikn\sigma})/(1-e^{2\pi ik\sigma})$ is zero for all $k$.  This is
only possible, if $n\sigma$ is an integer, i.e.\ if the cluster length
$n\sigma$ is an integer multiple of the wavelength. In that case, a
cluster potential could be invariant with respect to the cluster
position.  Since $\sigma<1$ [for $(N\!+\!1)$ particles and $N$
  potential wells, $\sigma<N/(N\!+\!1)$], it follows that $V_n(x)$ can
be constant only for diameters $\sigma$ equal to the rational numbers
\begin{equation}
\sigma_{m,n}=\frac{m}{n}\,,\hspace{1em}m=1,\ldots,n\!-\!1\,.
\end{equation}
However, for the factor $(1-e^{2\pi ikn\sigma})/(1-e^{2\pi ik\sigma})$
to be zero, $k\sigma$ must not be an integer, because otherwise
$(1-e^{2\pi ikn\sigma})/(1-e^{2\pi ik\sigma})\to n\ne0$ according to
L'Hospital's rule. This additional requirement of $k\sigma$ not being
an integer, cannot be satisfied for all $k$ if $\sigma=\sigma_{m,n}$.
It implies that the $c_k$ must be zero for those $k$, where
$k\sigma_{k,n}=km/n$ is an integer.

To specify the corresponding values $k$ for a given $\sigma_{m,n}$
(given $m$ and $n$), we set $m=d(m,n)m'$ and $n=d(m,n)n'$, where
$d(m,n)$ is the greatest common divisor of $m$ and $n$, and
accordingly $m'$ and $n'$ are coprime.  Then $km/n=km'/n'$ must be an
integer, which implies that $k$ is an integer multiple of
$n'=n/d(m,n)$. We therefore have to require
\begin{equation}
c_k=0\hspace{1em}\mbox{for}\hspace{0.3em}k=j\frac{n}{d(m,n)}\,,\hspace{0.6em} j=1,2,\ldots
\label{eq:cm-zeros}
\end{equation}
To summarize, for $V_n(x)$ to be constant, $\sigma$ has to be a
rational number. If this rational number is $m/n$, then all $c_k$ must
be zero for $k$ being an integer multiple of $n/d(m,n)$.

This sounds complicated, but the conditions can be satisfied for quite
general periodic potentials. Let us first consider a case, where the
periodic potential is represented by a finite number of Fourier
coefficients, i.e.\ where only the first $k_{\rm max}$ of the $c_k$
are nonzero ($c_k=0$ for $k>k_{\rm max}$). Then
Eqs.~\eqref{eq:cm-zeros} are fulfilled if $n/d(m,n)>k_{\rm max}$. This
means that clusters of size $n$ formed by particles with diameter
$\sigma_{m,n}=m/n$ move barrier-free if $n/d(m,n)>k_{\rm max}$. In
particular, since $d(n\!-\!1,n)=1$, all $n$-clusters formed by
particles with $\sigma=(n\!-\!1)/n$ move barrier-free for $n>k_{\rm
  max}$.
 
For the cosine potential considered in the main text, $k_{\rm max}=1$,
and Eqs.~\eqref{eq:cm-zeros} are satisfied for all $n$ and $m$,
i.e.\ potential barriers for $n$-clusters vanish for all $\sigma=m/n$,
$n=2,3,\ldots$, and $m=1,\ldots,n\!-\!1$. This is in agreement with
Eq.~(2) of the main text.

A further example is the asymmetric periodic potential
\begin{equation}
V(x) =2 - \frac{1}{2} \sin(2\pi x)-\frac{1}{12} \sin(4\pi x)
\end{equation} 
often used in ratchet models \cite{Ryabov/etal:2016, Reimann:2002}.
For this potential $k_{\rm max}=2$, and the nonzero Fourier
coefficients are $c_0=2$, $c_1=i/4$, and $c_2=i/24$.  Accordingly, all
$n$-clusters formed by particles with diameter $m/n$ and with
$n/d(m,n)>2$ move barrier-free: 3-clusters if $\sigma=1/3$ or 2/3,
4-clusters if $\sigma=1/4$ or 3/4, etc.\ Note that for a 4-cluster
formed by particles with diameter $2/4=1/2$, $d(m,n)=d(2,4)=2$,
$n/d(m,n)=2$, i.e.\ for such a cluster the potential is not constant
because $c_2\ne0$.

As an example for a periodic potential with an infinite number of
nonzero Fourier coefficients, let us consider the periodically
continued piecewise linear potential ($0<x_0<1$)
\begin{equation}
V(x)=\left\{\begin{array}{c@{\hspace{1em}}c}
\displaystyle V_0\, \frac{x}{x_0}\,, & 0\le x\le x_0\,,\\[2ex]
\displaystyle V_0\,\frac{1-x}{1-x_0}\,, & x_0\le x\le 1\,.
\end{array}\right.
\end{equation}
The Fourier coefficients for this potential are
\begin{equation}
c_k=-\frac{V_0}{4\pi^2 k^2}\frac{1-e^{-2\pi i kx_0}}{x_0(1-x_0)}\,.
\label{eq:ck-linear-potential}
\end{equation}
Barrier-free motions of clusters can be obtained if $x_0$ is a rational number, $x_0=q/p$ with $q,p\in\mathbb{N}$, with
$q<p$ coprime. Clusters for which $V_n(x)$ is constant then
are formed by an integer multiple of $p$ particles with diameter $q/n$, i.e.\ 
the possible $\sigma_{m,n}$ are given by $n=lp$ with $l\in\mathbb{N}$
and $m=q$. Since $x_0=q/p=ml/n$, the Fourier coefficients $c_k$ with $k=jn/d(n,m)$ 
in Eq.~\eqref{eq:ck-linear-potential} indeed vanish:
\begin{equation*}
\exp(-2\pi i kx_0)=\exp(-2\pi i \frac{n}{d(m,n)}j\frac{m}{n}l)=\exp(-2\pi i \frac{m}{d(m,n)}jl)=1\,.
\end{equation*}

\section{Soliton velocity and soliton-induced current}
\label{sec:soliton-velocities}
From the analysis of the deterministic motion in the zero-noise limit,
we find the conditions \eqref{eq:cond-fragmentation},
\eqref{eq:cond-nofragmentation} on the external forces to yield
solitons that are either of type $(n\!+\!1)$-$n$ or of type
$n$-$(n\!-\!1)$-$n$-$(n\!+\!1)$. Knowing this, we can calculate the
velocity of the solitons. This is exemplified here for the
$(n\!+\!1)$-$n$-solitons.

Consider a state of an $(n\!+\!1)$-$n$-soliton, where an
$(n\!+\!1)$-cluster is moving with particles having coordinates
$x_0(t), x_0(t)+\sigma,\ldots, x_0(t)+n\sigma$.  We fix the time
origin by saying that the $(n\!+\!1)$-cluster started moving at time
$t=0$ at an initial position $x_0(0)=x_0^{\rm ini}$.  The propagation
of the $(n\!+\!1)$-cluster terminates at a final position $x_0^{\rm
  fin}$, when the first particle detaches from the $n$-particles to
the right. According to the condition \eqref{eq:cond-fragmentation},
this happens if
\begin{equation*}
f+\pi U_1 \sin(2\pi x_0^{\rm fin}) = f+\frac{\pi U_1}{n} \sum_{k=1}^n \sin[2\pi(x_0^{\rm fin} + k\sigma)]
=f+\frac{\pi U_1\sin(n\pi\sigma)}{n\sin(\pi\sigma)}\sin\left[2\pi\left(x_0^{\rm fin}+\frac{n\!+\!1}{2}\sigma\right)\right]\,,
\label{eq:x0-at-separation}
\end{equation*}
yielding
\begin{equation}
\sin(2\pi x_0^{\rm fin})=\frac{\sin(n\pi\sigma)}{n\sin(\pi\sigma)}\sin\left[2\pi\left(x_0^{\rm fin}+\frac{n\!+\!1}{2}\sigma\right)\right]\,.
\end{equation}
One solution of this equation is
\begin{equation}
x_0^{\rm fin} = \frac{1}{2}+\frac{1}{2\pi}\arccot{\left|\frac{n\sin(\pi\sigma)}{\sin(\pi n \sigma) \sin[\pi(n\!+\!1)\sigma]}- \cot[\pi(n\!+\!1)\sigma] \right|}\,.
\end{equation}
Any $x_0^{\rm fin}+m$, $m=0,\ldots, (N\!-\!1)$, is a solution also,
i.e.\ when the distance of the position from $x_0^{\rm fin}$ equals an
integer multiple of the wavelength.

The final position of the $(n\!+\!1)$-cluster gives the initial
position $x_1^{\rm ini}$ for the $n$-cluster,
\begin{equation}
x_1^{\rm ini}=x_0^{\rm fin}+\sigma\,.
\end{equation}
The motion of the $n$-cluster terminates at a position $x_1^{\rm
  fin}$, when it attaches to the next (nearly) resting particle in the
system.  Then the motion of a new $(n\!+\!1)$-cluster starts, with a
position $x_0'$ of its first particle equal to $x_1^{\rm fin}$ [as the
  first particle of the attaching $n$-cluster becomes the first
  particle of the new $(n\!+\!1)$-cluster]. The $x_0'$ must be exactly
one wavelength apart from $x_0^{\rm ini}$, because the soliton motion
is periodically repeating itself after one wavelength. Accordingly,
\begin{equation}
x_1^{\rm fin}=x_0^{\rm ini}+1\,,
\end{equation}
where $x_0^{\rm ini}$ is a position of a resting particle, i.e.\ any position 
\begin{equation}
x_m^{\rm rest}=\frac{1}{2} + \frac{1}{2\pi} \arcsin(\frac{f}{\pi U_1}) +m\,,\hspace{1em} m=0,\ldots, N-1\,.
\end{equation}
of mechanical equilibrium in the tilted potential $[U(x)-fx]$.
We choose $m=0$, i.e.\
\begin{equation}
x_0^{\rm ini}=x_0^{\rm rest}=\frac{1}{2} + \frac{1}{2\pi} \arcsin(\frac{f}{\pi U_1})\,.
\end{equation}

\begin{figure}[t!]
\includegraphics[width=0.5\textwidth]{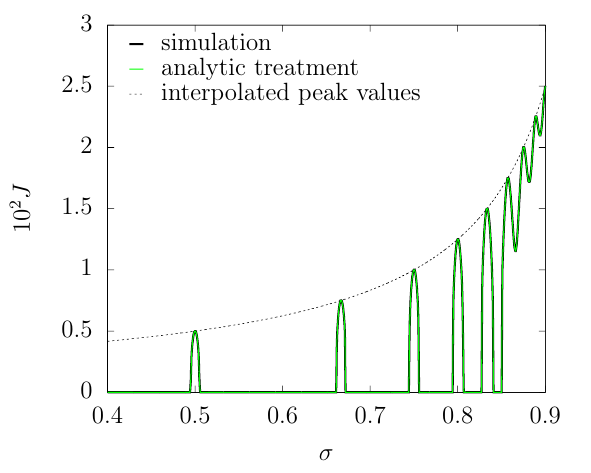}
\caption{Analytically calculated soliton-induced current
  [Eq.~\eqref{eq:Jsol}] as a function of particle diameter $\sigma$ in
  comparison with simulated data for $f=0.05$. The dotted line refers
  to an interpolation of the peak currents given in Eq.~(6) of the
  main text.}
\label{fig:soliton-velocity}
\end{figure}

The velocities $v_n$ and $v_{n+1}$ of the $n$ and $(n\!+\!1)$-cluster
are given by the mobility $\mu=1$ times the mean forces $\bar F_n$ and
$\bar F_{n+1}$ acting on the cluster. These mean forces are ($k=n,
n\!+\!1$)
\begin{equation}
\bar F_k(x)=\frac{1}{k}\sum_{i=0}^{k-1}F(x_i)=\frac{1}{k}\sum_{i=0}^{k-1} \bigl(f+\pi U_1\sin[2\pi (x+i\sigma)]\bigr)
=f+\frac{\pi U_1}{k}\frac{\sin(\pi k\sigma)}{\sin(\pi\sigma)}\sin[\pi(2x+(k-1)\sigma)]\,.
\end{equation}
The times $\tau_n$ and $\tau_{n+1}$ for the motion of the $n$ and
$(n\!+\!1)$-cluster in the intervals $[x_0^{\rm ini}, x_0^{\rm fin}[$
    and $[x_1^{\rm ini}, x_1^{\rm fin}[$ then are
\begin{subequations}
\begin{align}
\tau_n&=\int_{x_1^{\rm ini}}^{x_1^{\rm fin}}\hspace{-0.3em}\frac{\dd x}{\bar F_n(x)}\,,\\[1ex]
\tau_{n+1}&=\int_{x_0^{\rm ini}}^{x_0^{\rm fin}}\hspace{-0.3em}\frac{\dd x}{\bar F_{n+1}(x)}\,.
\end{align}
\end{subequations}
The time period of the soliton is $\tau=\tau_n+\tau_{n+1}$ and its
mean velocity $\bar v_{\rm sol}=1/\tau$. The current generated by the
soliton is
\begin{equation}
J=\frac{\bar v_{\rm sol}}{L}=\frac{1}{L(\tau_n+\tau_{n+1})}.
\label{eq:Jsol}
\end{equation}
This analytical result for the current is in excellent agreement with
simulated data, see Fig.~\ref{fig:soliton-velocity}.


%

\end{document}